\documentclass[superscriptaddress,reprint,amsmath,amssymb,aps,prl]{revtex4-2}
\usepackage{natbib}

\usepackage{rotating}

\usepackage{natbib}
\usepackage{amsmath,amssymb}
\usepackage{graphicx}
\usepackage{dcolumn}
\usepackage{bm}
\usepackage{xcolor} 

\newcommand{\beq}{\begin{equation}}
\newcommand{\eeq}{\end{equation}}

\newcommand{\duJS}{{\delta v_\lambda}}

\newcommand{\dperp}{{D}_{\perp \perp}}
\newcommand{\dpar}{{D}_{\parallel \parallel}}
\newcommand{\dcross}{{D}_{\parallel \perp}}

\newcommand{\dperpsh}{{D}_{\perp \perp}^{\mathrm{SH}}}
\newcommand{\dperprms}{{D}_{\perp \perp}^{\mathrm{SH, rms}}}
\newcommand{\dperpint}{{D}_{\perp \perp}^{\mathrm{SH, int}}}
\newcommand{\dperpintJ}{{D}_{\perp \perp}^{\mathrm{SH, int, J25}}}

\newcommand{\dQperp}{{dQ}_{\perp}}
\newcommand{\dQperpemp}{{dQ}_{\perp}^{\mathrm{emp}}}
\newcommand{\dQperpicw}{{dQ}_{\perp}^{\mathrm{ICW}}}
\newcommand{\dQperpsh}{{dQ}_{\perp}^{\mathrm{SH}}}
\newcommand{\dQperprms}{{dQ}_{\perp}^{\mathrm{SH, rms}}}
\newcommand{\dQperpint}{{dQ}_{\perp}^{\mathrm{SH, int}}}
\newcommand{\dQperpintJ}{{dQ}_{\perp}^{\mathrm{SH, int, J25}}}

\newcommand{\Qperpsh}{{Q}_{\perp}^{\mathrm{SH}}}

\newcommand{\kres}{{k}_{res}}
\newcommand{\wres}{{\omega}_{res}}

\newcommand\alf{Alfv\'en}
\newcommand\alfic{Alfv\'enic}

\newcommand\kms{\mathrm{km/s}}

\begin{document}

\title{Direct Measurement of Diffusion Coefficients: Evidence for Diffusive Stochastic Heating in Collisionless Plasmas}


\author{Tamar Ervin}
\email{tamarervin@berkeley.edu}
\affiliation{Department of Physics, University of California, Berkeley, CA 94720-7300, USA}
\affiliation{Space Sciences Laboratory, University of California, Berkeley, CA 94720-7450, USA}

\author{Trevor A. Bowen}
\affiliation{Space Sciences Laboratory, University of California, Berkeley, CA 94720-7450, USA}

\author{Alfred Mallet}
\affiliation{Space Sciences Laboratory, University of California, Berkeley, CA 94720-7450, USA}

\author{Philip A. Isenberg}
\affiliation{Department of Physics \& Astronomy, University of New Hampshire, Durham, NH 03824, USA}

\author{Kristopher G. Klein}
\affiliation{Lunar and Planetary Laboratory, University of Arizona, Tucson, AZ}

\author{Stuart D. Bale}
\affiliation{Department of Physics, University of California, Berkeley, CA 94720-7300, USA}
\affiliation{Space Sciences Laboratory, University of California, Berkeley, CA 94720-7450, USA}
\affiliation{The Blackett Laboratory, Imperial College London, London, SW7 2AZ, UK}

\author{Benjamin D. G. Chandran}
\affiliation{Department of Physics \& Astronomy, University of New Hampshire, Durham, NH 03824, USA}

\author{Roberto Livi}
\affiliation{Space Sciences Laboratory, University of California, Berkeley, CA 94720-7450, USA}

\author{Ali Rahmati}
\affiliation{Space Sciences Laboratory, University of California, Berkeley, CA 94720-7450, USA}

\author{Davin E. Larson}
\affiliation{Space Sciences Laboratory, University of California, Berkeley, CA 94720-7450, USA}

\begin{abstract} 
Open questions in collisionless plasma dissipation can be addressed using space-based observations in different astrophysical environments, with implications for both astrophysical and laboratory plasma systems. 
We study a low-$\beta$, highly imbalanced, sub-Alfv\'enic stream observed by Parker Solar Probe (PSP) to identify and distinguish between signatures of stochastic heating (SH) and resonant heating (RH) by parallel ion cyclotron waves (ICWs). 
Prior work studying this stream \citep{Bowen-2025prl} showed that the SH rate, accounting for intermittency, matched the amplitude of the local energy transfer (LET) rate while the RH rate did not. This comparison relied on a number of assumptions regarding the nature of the diffusive process, and the calculation of the LET rate.
We introduce a novel technique of inverting the proton guiding center equation to empirically measure velocity-space diffusion coefficients using three-dimensional proton velocity distribution functions (VDFs), from the ion electrostatic analyzer (SPANi) on PSP. 
Measured diffusion coefficients are used to determine phase-space heating rates, leading to a calculation of a fully kinetic heating rate independent of assumptions made in prior work.
We show that scale-dependent analytic expressions for SH via non-coherent fluctuations match the empirical measurements from PSP data, provided that we account for intermittency in the heating calculation. In contrast, the derived heating rates for SH that accounts for the effects of the helicity barrier, and heating rates for RH via $\parallel$-ICWs do not peak in the same region of velocity-space as the empirical measurements, nor reach the required magnitude. 
Our approach provides novel methodology to uniquely identify and constrain heating processes in collisionless plasmas, and shows evidence of a Fokker-Planck like diffusive process in the near-Sun solar wind.

\end{abstract}

\maketitle 

\paragraph{Introduction} \label{sec:intro}
As the solar wind expands outwards from the Sun's corona and through the inner heliosphere, it cools more slowly than would be expected via adiabatic expansion \citep{Chew-1956}. Double adiabatic theory, which describes the evolution of the distribution function in the presence of a background magnetic field, predicts strong parallel temperature anisotropies ($T_\parallel / T_\perp \gg 1$, where $T_\perp$ and $T_\parallel$ are defined relative to the background field) as the wind evolves. However, observations of proton velocity distribution functions (VDFs) show large perpendicular temperature anisotropies ($T_\perp / T_\parallel > 1$), especially near the Sun \citep[e.g.][]{Marsch-1982VDF, Huang-2025}. There must be additional heating as the wind evolves to explain the decades of observations of solar wind temperatures as a function of heliocentric distance \citep[e.g.][]{Richardson-1995}, and the heating mechanisms at work must preferentially accelerate particles perpendicular to the background field to explain the observations of large perpendicular anisotropies \citep{Matteini-2007}. 

Various heating mechanisms have been proposed to explain these observational signatures: non-resonant (stochastic) processes heating the distribution via uncorrelated kicks due to fluctuations in the fields \citep[SH;][]{Chandran-2010, Klein-2016, Mallet-2025SH, Bowen-2025prl}, resonant heating \citep[RH;][]{Cranmner-2000, Bowen-2022, Bowen-2024}, magnetic reconnection \citep{Vech-2018}, and others \cite{Verscharen-2019}. Recent work has suggested that the heating mechanisms may be sensitive to the level of balance between counter-propagating {\alfic} fluctuations \citep{Johnston-2025, Zhang-2025}, leading to a helicity barrier (HB) in the wave number spectrum that prevents strongly imbalanced turbulence from cascading to small-scales \citep{Meyrand-2021, Squire-2022}.

Observational work has directly connected levels of $\parallel$-cyclotron resonant damping with turbulent heating in the super-{\alfic} solar wind \cite{Bowen-2022, Bowen-2024}. Additionally, a numerical study of forced imbalanced turbulence found that RH dominates as the HB prevents the cascade of fluctuations to the smallest perpendicular scales resulting in oblique-cyclotron RH that drives $\parallel$ propagating waves via an {\alf} ion cyclotron wave instability \citep{Squire-2022}. In the solar wind, such $\parallel$ propagating waves can propagate outward from the Sun and may result in net heating via cyclotron damping, seen both observationally \citep{Shankarappa-2024} and in simulations. However in the sub-{\alfic} solar wind, \citet{Bowen-2025prl} (B25 moving forward) has shown that cyclotron resonant heating does not provide the required heating to match observed cascade rates, but rather suggest SH as a viable mechanism to provide the requisite energy to heat the sub-{\alfic} wind, provided intermittent fluctuations are accounted for. This previous work assumed perpendicular diffusion via intermittent structures to calculate the SH rate, without any evidence that this process was actually occurring. Simulations show evidence for SH that can perpendicularly heat the distribution as it evolves through the inner heliosphere \cite{Chandran-2010, Klein-2016, Arzamasskiy-2019, Cerri-2021}. Observational work has estimated coefficients associated with the SH rate at different heliocentric distances \citep{Bourouaine-2013} and calculated dissipation rates throughout the heliosphere \citep{Martinovic-2019, Martinovic-2020}. These works were limited in their heating rate calculations to computation from scalar moments and spectra rather than a calculation of the kinetic phase-space heating rate from the full distribution function. 

Each of these heating mechanisms is associated with a diffusive process and a distinctive set of diffusion coefficients. Here, we use a fully kinetic framework to make an empirical measurement of diffusion coefficients and heating rates as a function of velocity-space from measured VDFs in a sub-{\alfic} solar wind stream observed by Parker Solar Probe \citep[PSP;][]{Fox-2016}. We directly compare our empirical heating rates with proposed heating rate expressions for SH \cite{Chandran-2010, Klein-2016, Isenberg-2019, Johnston-2025, Mallet-2025SH} and RH via $\parallel$ ion cyclotron waves (ICWs) \citep{Isenberg-2007}. Our results show that scale-dependent analytic expressions for SH rates, including intermittent effects, match empirical measurements in the sub-{\alfic} solar wind. In contrast, a recent SH expression that assumes heating to be maximized at a particular scale to account for the effects of a HB \citep{Johnston-2025}, peaks in the incorrect region in velocity-space to match our empirical measurements. Similarly, expressions for RH via $\parallel$-ICWs do not match observations. These comparisons allow us to test the validity of assumptions made in prior works (e.g. B25), as our methodology enables direct computation of diffusion coefficients and heating rates from empirical observation of the VDF evolution, without assumption of a specific mechanism. This work provides a novel method to calculate phase-space diffusion coefficients and fully-kinetic heating rates for identification of dominant heating mechanisms at work in collisionless plasmas and constraining coefficients associated with theoretical expressions with wide applicability in both observations and simulations of space and laboratory plasma environments.



\begin{figure*}
    \centering
    \includegraphics[width=\linewidth]{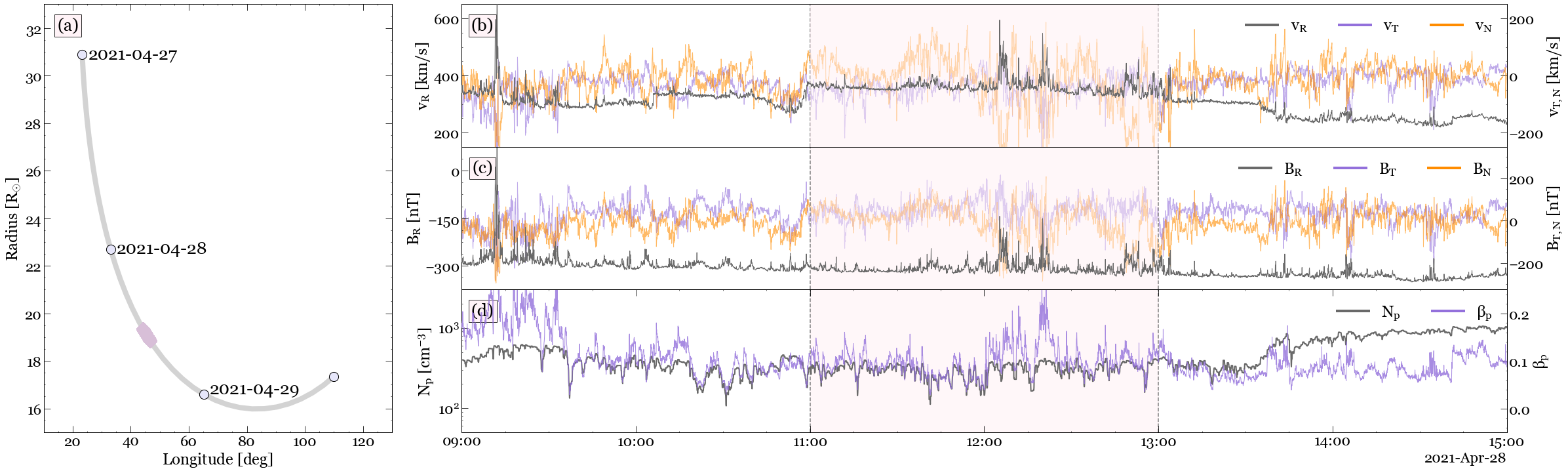}
    \caption{An overview of the time period of interest in this study. (a) PSP trajectory (longitude vs. radius). (b) Velocity measurements from PSP/SPANi. (c) Fluxgate magnetic field measurements from PSP/FIELDS. (d) Proton density measurements from PSP/SPANi (partial moment). We take the maximum value across a rolling window of 10 observations ($\sim 35$ seconds) to account for density dropouts due to FOV effects. Proton plasma $\beta$ is shown in purple, $\beta_p = n k_B T / (B^2 / (2 \mu_0))$, calculated from the PSP/SPANi density shown in (d). The highlighted portion is our time period of interest.}
    \label{fig:data}
\end{figure*}

\paragraph{Data} \label{sec:data}
PSP provides measurements of the electromagnetic fields and charged particles in the inner heliosphere allowing us to measure and constrain the heating and acceleration mechanisms powering the solar wind. We focus on a portion of the previously studied first sub-{\alfic} stream from PSP Encounter 8 \citep[e.g.][]{Kasper-2021, Zank-2022, Zhao-2022, Bowen-2025prl}, where PSP enters the magnetically dominated corona and the spacecraft is flying radially towards the Sun with little change in Carrington longitude. We study the two-hour interval from 11:00 to 13:00 on April 28, 2021 (highlighted in Fig.~\ref{fig:data}) when we estimate the source connectivity \citep[following][]{Ervin-2024c} to be stable, even with varied model input parameters, and connected to a single negative polarity coronal hole. This allows for study of the evolution sub-{\alfic} stream from a stable source. This is a low-$\beta$ ($\beta \sim 0.1$), highly imbalanced ($\sigma_C \sim 0.9$) stream, indicating applicability of SH and RH mechanisms.

We use observations of proton VDFs from the Ion Solar Probe ANalyzers \citep[SPANi;][]{Livi-2022, Kasper-2016} on PSP which provides 3d measurements of the velocity distribution at a cadence of $\sim$3.5 s. This period has well-resolved proton distribution functions where the bulk of the distribution falls within the instrument's field-of-view (FOV; see End Matter). Fig.~\ref{fig:data}(b) shows velocity measurements (from SPANi partial moments) in the Radial-Tangential-Normal (RTN) coordinate system and Fig.~\ref{fig:data}(d) shows the partial proton density moment from SPANi and proton plasma beta ($\beta_p$). High-resolution magnetic field measurements using the merged fluxgate and search coil data product \cite[SCaM;][]{Bowen-2020-SCaM} from the FIELDS suite \citep{Bale-2016} are used to make measurements of the in situ fluctuations at a high frequency. Fig.~\ref{fig:data}(c) shows the fluxgate magnetic field measurements in RTN coordinates.  

\paragraph{Empirical Calculation of Diffusion Coefficients}\label{sec:diffusion}

We use the proton guiding center equation (Eq.~\ref{eqn:guiding-center-no-radial}), which describes the steady-state evolution of a gyrotropic particle distribution \citep{Kulsrud-1983, Isenberg-1997}, to calculate diffusion coefficients in the sub-{\alfic} wind stream:

\begin{multline} \label{eqn:guiding-center-no-radial}
        (U + v_\parallel) \frac{\partial f}{\partial r} + \frac{v_\perp}{2} \frac{d \ln{A}}{dr} \left(v_\perp \frac{\partial f}{\partial v_\parallel} - (U + v_\parallel) \frac{\partial f}{\partial v_\perp} \right) =  \\
    \frac{\partial}{\partial v_\parallel} \left( D_{\parallel \parallel} \frac{\partial f}{\partial v_\parallel} + D_{\parallel \perp} \frac{\partial f}{\partial v_\perp} \right) \\
    + \frac{1}{v_\perp} \frac{\partial}{\partial v_\perp} \left[v_\perp \left(D_{\parallel \perp} \frac{\partial f}{\partial v_\parallel} + D_{\perp \perp} \frac{\partial f}{\partial v_\perp} \right) \right] .
\end{multline} The equation is written in the bulk plasma reference frame, moving outward at the radial flow speed, $U(r)$, and phase-space variables ($v_\perp, v_\parallel$) are defined relative to this frame. The left-hand side contains the advection term and double-adiabatic (CGL) terms respectively, while the right-hand side accounts for diffusion by small-scale fluctuations. $D_{ii}$ are the diffusion coefficients, $r$ is heliocentric radial position and $A(r)$ is the flux-tube area as a function of radial distance. For this essentially local analysis, we ignore large-scale radial forces such as gravity and the ambipolar electric field. We define the flux-tube area as \citep{Isenberg-2019}: $ A(r) \sim \frac{R^6}{R^4+6}$ where $R = r / R_\odot$.

All partial derivatives are taken using centered finite differences to conserve total particle density in the distribution functions \citep{Richardson-1955-finite, Zhou-1993, Vasquez-2020, Wole-2021}. To take these centered finite differences of the distribution functions (e.g. $\partial f / \partial v_i$), the VDFs must be fit and interpolated onto a uniformly spaced grid, in velocity-space. We use radial basis functions (RBF) to fit the measured SPANi VDFs \citep{Bowen-2024apjl, Bowen-2023-ursi} and interpolate them onto a uniformly spaced 5 ${\kms}$ grid. The RBF method better captures the non-Maxwellian features of the distribution functions produced by collisionless processes, in comparison to typical bi-Maxwellian fitting methods \citep{Bowen-2024}, and has been used extensively in the solar wind to study heating \citep[e.g.][]{Bowen-2024apjl}. 

With the calculation of gradients both in phase space ($\partial f / \partial v_i$) and between spacecraft observations ($\partial f / \partial r$) and the in-situ measurements of the bulk flow $U$, the left-hand side of Eq.~(\ref{eqn:guiding-center-no-radial}) and all terms on the RHS other than $D_{ii}$ can be determined. We require a set of nine data points in ($r, v_\perp, v_\parallel$) space to create a fully constrained system of equations. We invert the system using singular value decomposition \citep[SVD;][]{Branham-1990} to solve for ${\dperp}$, ${\dpar}$, and ${\dcross}$ as a function of $(v_\perp, v_\parallel$), assuming cylindrical symmetry ($D_{\perp \parallel} = D_{\parallel \perp}$). This method produces 2d measurements of $D_{ii} (v_\perp, v_\parallel)$ over time that can be used to compute heating rates and compared to theoretical diffusion coefficients. As nine VDF data points are required to constrain this calculation, we use a rolling window across the interval to get 116 measurements of the diffusion coefficients over the two-hour interval. The Supplementary Material \citep{Ervin-2026-prl-sm} includes a full derivation of the inversion process including the offset between the VDF grid and the $D_{ii}$ grid leading to the necessity for nine observation points to constrain the equation. Here, we focus on ${\dperp}(v_\perp, v_\parallel)$ and derived quantities (e.g. $\dQperp$). Future work will examine ${\dpar}$ and ${\dcross}$.

\begin{figure*}
    \centering
    \includegraphics[width=\linewidth]{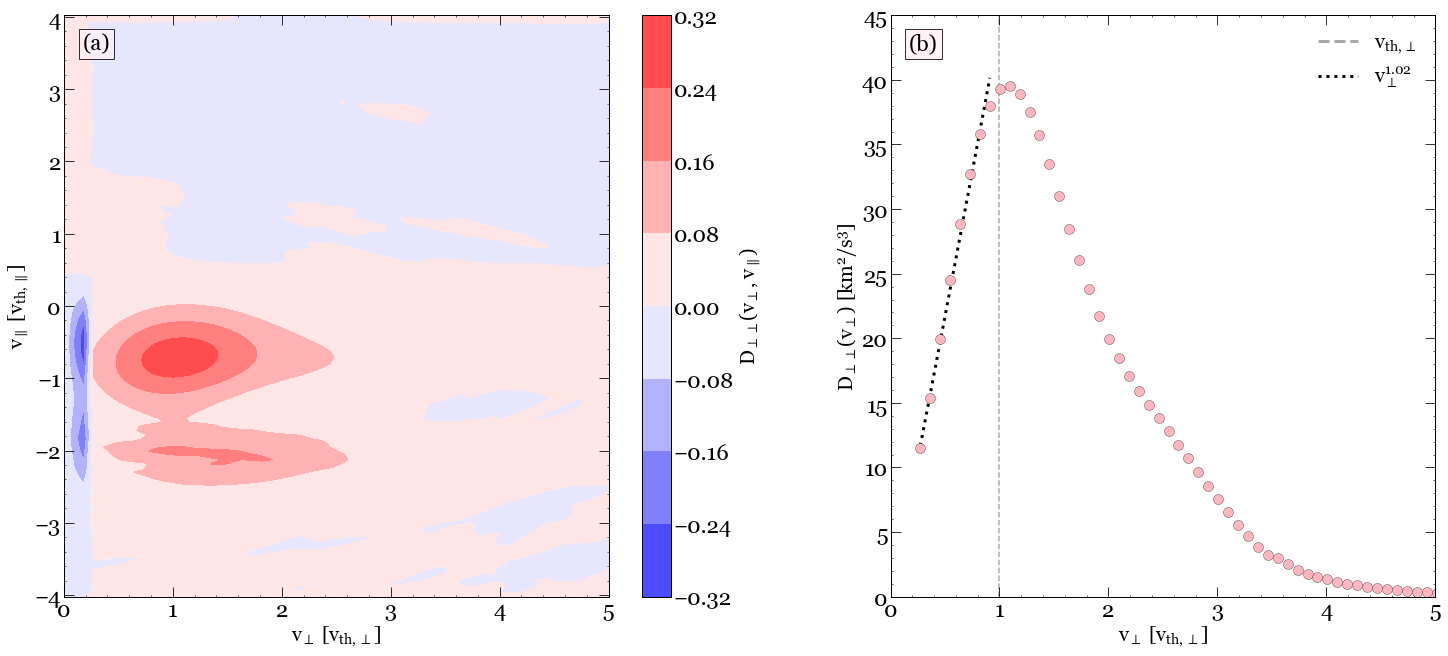}
    \caption{Example of the mean perpendicular diffusion coefficients ($\dperp$) calculated over the time period of interest. (a) Mean 2d $\dperp$ over time period of interest. (b) Mean 1d ${\dperp}(v_\perp) = \int {\dperp}(v_\perp, v_\parallel) d v_\parallel$ over period. The integration of each calculated 2d coefficient is preformed prior to taking the mean. The dotted black line shows the fit to the portion below the average $v_{th, \perp}$ (vertical dashed gray line) for the interval. }
    \label{fig:diffusion}
\end{figure*}

In Fig.~\ref{fig:diffusion}(a), we show the mean (over the full interval) 2d perpendicular diffusion coefficient as a function of phase space. $\dperp(v_\perp, v_\parallel)$ peaks near $v_{th, \perp}$, $v_{th, \parallel}$ and $2 v_{th, \parallel}$. We numerically integrate $\dperp(v_\perp, v_\parallel)$ to calculate a 1d velocity diffusion coefficient, ${\dperp}(v_\perp) = \int_{- \infty}^\infty {\dperp}(v_\perp, v_\parallel) d v_\parallel$. We calculate ${\dperp}(v_\perp)$ associated with each $\dperp(v_\perp, v_\parallel)$ measurement as shown in Fig.~\ref{fig:diffusion}(b). A fit of $\dperp(v_\perp)$ where  $v_\perp < \langle v_{th, \perp} \rangle$ shows $\dperp \propto v_\perp^{1.02}$. The diffusion coefficient peaks near $v_{th, \perp}$ as to be expected for SH \citep{Mallet-2025SH}. 

With these empirical measurements of ${\dperp}(v_\perp)$ we directly compare to proposed analytic expressions for SH \citep{Chandran-2010, Klein-2016}, RH via $\parallel$-ICWs \cite{Bowen-2024apjl}, and a formulation that accounts for the effects of the HB \cite{Johnston-2025}. We calculate a perpendicular heating rate from $\dperp (v_\perp)$ through integration by parts \citep{Klein-2016, Cerri-2021}:
\begin{equation} \label{eqn:perp-heating}
    Q_\perp = \int dQ_\perp = -2 \pi m_i \int_0^\infty v_\perp^2 \dperp (v_\perp) \frac{\partial f_\perp}{\partial v_\perp} dv_\perp 
\end{equation} where $f_\perp = \int f(v_\perp, v_\parallel) d v_\parallel$ is the reduced distribution function from the RBF fits to the observed VDFs.

\paragraph{Comparison with Stochastic and Cyclotron Resonant Heating} \label{sec:sh}

Analytic expressions for SH have been proposed by a variety of authors \citep[e.g.][]{Chandran-2010, Klein-2016, Johnston-2025, Mallet-2025SH}, referring to a process whereby large-amplitude fluctuations can disrupt a particle's gyromotion, breaking the conservation of the magnetic moment, leading to perpendicular heating of the particle distribution. SH typically depends on a \lq{}stochasticity parameter\rq{} ($\varepsilon = \delta v / v$) that describes the efficiency of the process. There are various definitions of $\varepsilon$ depending on where theory predicts heating to be most efficient, as well as methods that use scale-dependent versions of $\varepsilon$ \cite{Klein-2016, Cerri-2021, Bowen-2025prl}. 

Regardless of the definition of $\delta v$ and the corresponding stochasticity parameter $\varepsilon$, these SH formulations typically provide an analytic expression for $\dperpsh$ in low-$\beta$ {\alfic} turbulence \cite{Chandran-2010}:
\begin{equation} \label{eqn:sh-diffusion}
    \dperpsh = c_1 \frac{\Omega_i}{v_i} \delta v^3 e^{-c_2 / \varepsilon}
\end{equation} where differences between formulations lie in the definitions of $\delta v$ and $\varepsilon$ (see End Matter). $c_1$ and $c_2$ are (order unity) coefficients that account for assumptions in the physics of these models or can be fit parameters \citep{Martinovic-2019}. We use $c_1 = 0.75$ and $c_2 = 0.34$, determined via test-particle simulations \citep{Chandran-2010}. $\dperpsh$ is used in Eq.~\ref{eqn:perp-heating} to determine $\Qperpsh$. $\dperpsh$ includes an exponential term ($\mathrm{e^{\frac{-c_2}{\varepsilon}}}$) that accounts for the suppression of the heating where $\varepsilon$ is small. This suppression factor was added semi-empirically in many formulations and recent work has shown that this term holds through a derivation from first principles \citep{Mallet-2025SH}. 

We compare different proposed expressions for $\delta v$ in the SH diffusion coefficient: a scale-dependent formulation \citep{Klein-2016}, and a formulation using fluctuations at a particular scale to account for HB effects, $\delta v (v_\perp^m)$ the \citep[J25 moving forward;][]{Johnston-2025}. The versions of $\delta v$ are calculated from the PSP SCaM data (see End Matter). 

As $\dperpsh \propto \delta v^3$, large-amplitude, intermittent fluctuations can have a dramatic effect on the SH diffusion coefficient. We calculate $\dQperpsh$ with \lq{}rms\rq{} ($\dperprms$) and \lq{}intermittent\rq{} ($\dperpint$) velocity diffusion coefficients \citep{Bowen-2025prl}. $\dperprms$ is defined using the rms fluctuation amplitude $\delta v^{\mathrm{rms}}(v_\perp) = \sqrt{\langle \delta v(v_\perp, t)^2 \rangle}$ over the time interval, thus averaging out the effects of large-amplitude intermittent fluctuations. $\dperpint (v_\perp) = \langle |\dperp(v_\perp, t)| \rangle$ to account for intermittency. We produce a scale-dependent $\dperpsh(v_\perp)$ from which we can calculate $\dQperpsh(v_\perp)$ (Eq.~\ref{eqn:perp-heating}). We note that the J25 formulation relies on 
the fluctuations at a particle scale  ($\duJS(v_\perp^m)$) associated with the steepening of the turbulent power spectrum (see End Matter). Thus, $\dperpintJ$ is not a function of scale, but rather a single value at all $v_\perp$. The velocity-space dependence of the intermittent J25 heating rate, $\dQperpintJ$, comes from Eq.~\ref{eqn:perp-heating}.



\begin{figure}
    \centering
    \includegraphics[width=\linewidth]{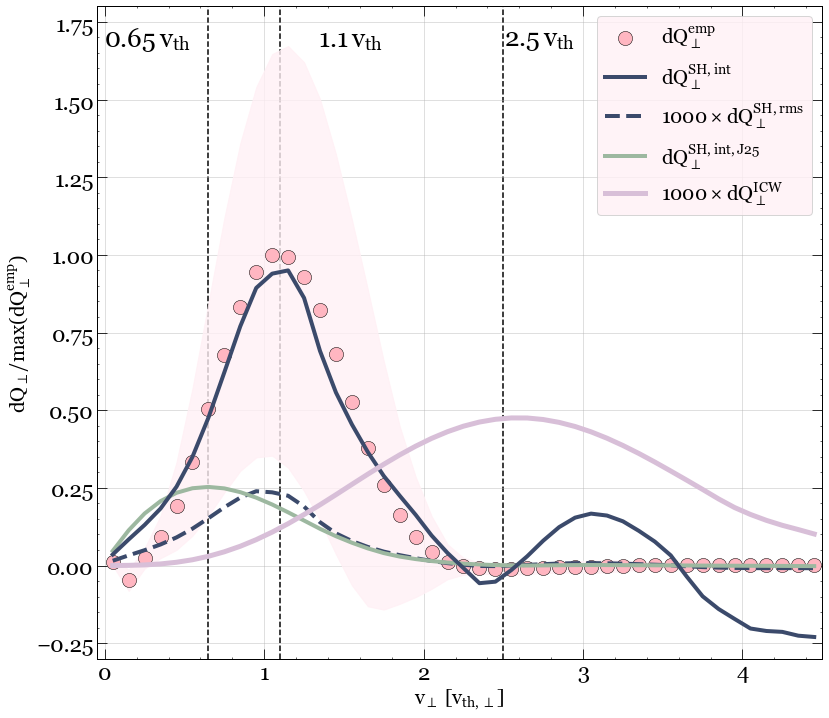}
    \caption{A comparison of the heating rates calculated from the empirical measurement (pink), `intermittent' SH (blue), `rms' SH (dashed blue), J25 method with intermittency (green), and $1000 \times$ RH via ICWs (purple). The versions of the SH expression use canonical coefficients from prior work: $c_1 = 0.75$ and $c_2 = 0.34$ \citep{Chandran-2010}. The dashed line at $0.65 v_{th, \perp}$, $1.1 v_{th, \perp}$, $2.5 v_{th, \perp}$ shows the location of the peak in $\dQperpintJ$, $\dQperpemp$, and $\dQperpicw$ respectively. The pink shaded region shows the error from the empirical calculation. All measurements are normalized to the maximum value in the $\dQperpemp$ calculation.
    }
    \label{fig:sh-comparison}
\end{figure}

Fig.~\ref{fig:sh-comparison} compares $\dQperpemp$ with the three SH formulations and the perpendicular heating rate for RH via $\parallel$-ICWs ($\dQperpicw$; see End Matter). We find that $\dQperpemp$ peaks near $1.1 v_{th, \perp}$, consistent with results from simulations in low-$\beta$ plasmas \citep{Cerri-2021, Arzamasskiy-2019} and theoretical curves \citep{Mallet-2025SH}, then rapidly drops off after. While both the $\dQperprms$ and $\dQperpint$ curves peak near $1.1 v_{th, \perp}$, only the case that includes intermittency reaches the required heating magnitude. We see a secondary peak in the intermittent SH rate at $\sim 3 v_{th, \perp}$ that we believe could be attributed to a breakdown in the assumptions of the SH model at these perpendicular speeds, where fluctuations cause a smooth rather than chaotic drift to the ion’s guiding center motion. The J25 heating rate peaks at $\sim 0.65 v_{th, \perp}$ and is too small to reach the required magnitude. $\dQperpicw$ is significantly smaller, and peaks at $2.5 \; v_{th, \perp}$. 


\paragraph{Discussion}

The interplay between turbulent dissipation and collisionless heating is important in a range of plasma systems \citep[e.g.][]{Coleman-1968, Quataert-1998, Leamon-1998, Schekochihin-2007, Zhuravleva-2014}. This letter highlights a novel method to empirically measure diffusion coefficients that can serve as a fundamental constraint on collisionless heating mechanisms. We have applied this method in the sub-Alfv\'enic solar wind to constrain SH, which various theoretical \citep{Chandran-2010, Klein-2016, Mallet-2025SH}, simulation \citep{Arzamasskiy-2019, Cerri-2021}, and observational \citep{Martinovic-2019, Martinovic-2020, Bowen-2025prl} works have shown is viable in low-$\beta$ plasmas. Prior work calculating SH rates assumed a diffusive process without any evidence for one existing (e.g. B25). Our empirical measurement, with minimal assumptions, of a physically meaningful diffusion coefficient suggests the presence of a Fokker-Planck type process, underlying collisionless heating mechanisms that are responsible for plasma heating in the outer corona.

Our results show agreement of the scale-dependent analytic expression for the SH heating rate with empirical measurements, provided that intermittency, and thus the effect of infrequent, large-amplitude fluctuations, are accounted for, consistent with prior theoretical work \citep{Mallet-2019}. Our direct estimate of diffusive heating is not subject to a number of critical assumptions made by B25 (see End Matter). The general agreement between these two studies largely validates many of these assumptions, particularly in constraints on $c_2$, which is critical in understanding the efficiency and suppression of stochastic heating. We find our empirical measurement does not match with the recent formulation proposed by J25 to account for a variable level of balance in the turbulence, and thus effects of the helicity barrier \citep{Johnston-2025}, despite this stream being highly imbalanced. While the $c_1$ parameter could be increased by a factor of $\sim 4$ in $\dQperpintJ$ to reach the required heating magnitude, this heating rate peaks in the incorrect region in velocity-space and thus is not a viable description of heating in this stream. Future work could include fitting the empirically measured coefficients and heating rates to determine values of $c_1$ and $c_2$, and comparing these measurements during periods of balanced and imbalanced turbulence to understand the impacts of steepening in the turbulent spectrum.

During the time period of interest, PSP was traversing the outer corona in the sub-{\alfic} wind. This time period is ideal to study as the VDFs are well-resolved and we observe a radially-evolving plasma parcel, both necessary for the inversion method. These results thus provide direct insight into coronal heating processes and showcase the importance of SH in this regime. Future statistical work comparing sub-{\alfic} intervals with their super-{\alfic} counterparts could help determine if and where the transition from SH to RH, which has been reported at larger distances \citep{Bowen-2022}, takes place. \citet{Martinovic-2019, Martinovic-2020} quantified SH rates in the solar wind, not including intermittent effects, and found that the SH rate increases closer to the Sun. Understanding the origin of this transition and how it is modulated also remains an open question that could be addressed via the novel method described here.

Future work using this methodology to calculate solar wind heating rates across discontinuities, in super-{\alfic} streams, and in reconnection exhausts would allow us to understand and constrain the phase-space signatures of heating mechanisms in various regions of the heliosphere which can be used to drive global models \citep{Ervin-2026natcomms}. Similar calculations in different environments (e.g. planetary shocks, the high-$\beta$ plasma sheet) with varying dimensionless parameters (e.g. $\beta$, $m_A$, $\sigma_C$), have implications for our understanding of heating processes relevant to astrophysical plasmas in a range of environments that lack fully kinetic measurements, or cannot be directly measured by spacecraft. These results could be used to test theoretical predictions for dominant heating mechanisms based on fundamental plasma parameters \citep[e.g.][]{Howes-2024}. 

This methodology could be applied to multi-spacecraft constellations (e.g. the Magnetospheric Multiscale Mission) to calculate heating rates over spatial structures, such as Earth's bow shock, and compared with proposed heating mechanisms to understand the phase-space dependence of the observed heating. Future multi-spacecraft missions could utilize these methods and would require fewer time steps to calculate the spatial gradients necessary to constrain the calculation. Understanding the phase-space signatures of dissipation mechanisms and the evolution of the distribution function, can inform theories for entropy growth in collisionless systems \citep{Zhdankin-2022, Du-2020}. The field particle correlation technique \citep[FPC;][]{Klein-2016-FPC}, uses single point measurements of the electric field and particle distributions to directly measure local energy transfer rates between fields and particles and characterize the mechanisms enabling the transfer. Comparisons between our methods and the FPC technique when high fidelity electric field measurements are available could provide further constraints on velocity-space signatures of collisionless damping mechanisms.

Through this novel methodology for calculating heating rates as a function of velocity-space, we show the ability to observationally constrain which heating mechanisms operate in the solar wind, and which analytical formulations properly describe the velocity-space signatures, independently of assumptions made in many prior works \citep[e.g.][]{Bowen-2025prl, Martinovic-2019}. We show evidence for a Fokker-Planck type diffusive process in the sub-{\alfic} stream of study, and that scale-dependent SH rates, which include intermittency, match our empirical measurements both in magnitude, and in the region of phase space where these heating rates peak. This is a compelling constraint on turbulent heating and collisionless dissipation in the near-Sun environment. These methods provide a new pathway for identification of heating mechanisms in collisionless plasmas, and constraining coefficients associated with analytic models of heating mechanisms. 
 
\paragraph{Acknowledgments}
The FIELDS and SWEAP experiments on the Parker spacecraft was designed and developed under NASA contract NNN06AA01C. We acknowledge the NASA Parker Solar Probe Mission, the FIELDS team led by S. D. Bale, and the SWEAP team led by M. Stevens for use of data. 

TE acknowledges funding from The Chuck Lorre Family Foundation Big Bang Theory Graduate Fellowship. TAB acknowledges support from NASA Grant 80NSSC24K0272 through the HSR program. AM acknowledges support from NASA grant 80NSSC20K1284. KGK acknowledges partial support from NASA grant 80NSSC24K0171. BC acknowledges support from NASA Grant 80NSSC24K0171 and DOE grant DE-SC0026201.

The authors appreciate the feedback and suggestions given by the three anonymous referees, and their roles in improving the manuscript.
\bibliography{ms}{}

@PREAMBLE{
 "\providecommand{\noopsort}[1]{}" 
 # "\providecommand{\singleletter}[1]{#1}%" 
}

@string{apj  = {The Astrophysical Journal}}

@string{apjl = {The Astrophysical Journal Letters}}

@string{apjs = {The Astrophysical Journal Supplement Series}}

@string{grl  = {Geophysical Research Letters}}

@string{jgr  = {Journal of Geophysical Research}}

@string{ssr  = {Space Science Reviews}}

@misc{Ervin-2026-prl-sm,
 author = {{Ervin}, Tamar},
  note = {See Supplemental Material at https://journals.aps.org/prl/supplemental/10.1103/s4q4-v7rm for a derivation of the empirical calculation made from the spacecraft observations. It provides additional information for how the calculation was carried out, specifically: an overview of the discretization of the expression and implementation of centered finite differences.}
}

@ARTICLE{Quataert-1998,
       author = {{Quataert}, Eliot},
        title = "{Particle Heating by Alfv{\'e}nic Turbulence in Hot Accretion Flows}",
      journal = apj,
         year = 1998,
        month = jun,
       volume = {500},
       number = {2},
        pages = {978-991},
          doi = {10.1086/305770},
archivePrefix = {arXiv},
       eprint = {astro-ph/9710127},
 primaryClass = {astro-ph},
       adsurl = {https://ui.adsabs.harvard.edu/abs/1998ApJ...500..978Q}
}

@ARTICLE{Coleman-1968,
       author = {{Coleman}, Jr., Paul J.},
        title = "{Turbulence, Viscosity, and Dissipation in the Solar-Wind Plasma}",
      journal = apj,
         year = 1968,
        month = aug,
       volume = {153},
        pages = {371},
          doi = {10.1086/149674},
       adsurl = {https://ui.adsabs.harvard.edu/abs/1968ApJ...153..371C}
}

@ARTICLE{Du-2020,
       author = {{Du}, Senbei and {Zank}, Gary P. and {Li}, Xiaocan and {Guo}, Fan},
        title = "{Energy dissipation and entropy in collisionless plasma}",
      journal = {\pre},
         year = 2020,
        month = mar,
       volume = {101},
       number = {3},
          eid = {033208},
        pages = {033208},
          doi = {10.1103/PhysRevE.101.033208},
archivePrefix = {arXiv},
       eprint = {1911.08086},
 primaryClass = {physics.plasm-ph},
       adsurl = {https://ui.adsabs.harvard.edu/abs/2020PhRvE.101c3208D}
}

@ARTICLE{Zhdankin-2022,
       author = {{Zhdankin}, Vladimir},
        title = "{Generalized Entropy Production in Collisionless Plasma Flows and Turbulence}",
      journal = {Physical Review X},
         year = 2022,
        month = jul,
       volume = {12},
       number = {3},
          eid = {031011},
        pages = {031011},
          doi = {10.1103/PhysRevX.12.031011},
archivePrefix = {arXiv},
       eprint = {2110.07025},
 primaryClass = {astro-ph.HE},
       adsurl = {https://ui.adsabs.harvard.edu/abs/2022PhRvX..12c1011Z}
}

@ARTICLE{Leamon-1998,
       author = {{Leamon}, Robert J. and {Smith}, Charles W. and {Ness}, Norman F. and {Matthaeus}, William H. and {Wong}, Hung K.},
        title = "{Observational constraints on the dynamics of the interplanetary magnetic field dissipation range}",
      journal = jgr,
         year = 1998,
        month = mar,
       volume = {103},
       number = {A3},
        pages = {4775-4788},
          doi = {10.1029/97JA03394},
       adsurl = {https://ui.adsabs.harvard.edu/abs/1998JGR...103.4775L}
}

@ARTICLE{Zhuravleva-2014,
       author = {{Zhuravleva}, I. and {Churazov}, E. and {Schekochihin}, A.~A. and {Allen}, S.~W. and {Ar{\'e}valo}, P. and {Fabian}, A.~C. and {Forman}, W.~R. and {Sanders}, J.~S. and {Simionescu}, A. and {Sunyaev}, R. and {Vikhlinin}, A. and {Werner}, N.},
        title = "{Turbulent heating in galaxy clusters brightest in X-rays}",
      journal = {\nat},
         year = 2014,
        month = nov,
       volume = {515},
       number = {7525},
        pages = {85-87},
          doi = {10.1038/nature13830},
archivePrefix = {arXiv},
       eprint = {1410.6485},
 primaryClass = {astro-ph.HE},
       adsurl = {https://ui.adsabs.harvard.edu/abs/2014Natur.515...85Z}
}

@INCOLLECTION{Schekochihin-2007,
       author = {{Schekochihin}, Alexander A. and {Cowley}, Steven C.},
        title = "{Turbulence and Magnetic Fields in Astrophysical Plasmas}",
    booktitle = {Magnetohydrodynamics: Historical Evolution and Trends},
         year = 2007,
       editor = {{Molokov}, S. and {Moreau}, R. and {Moffatt}, H.~K.},
        pages = {85},
          doi = {10.48550/arXiv.astro-ph/0507686},
       adsurl = {https://ui.adsabs.harvard.edu/abs/2007mhet.book...85S}
}

@ARTICLE{Ervin-2026natcomms,
       author = {{Ervin}, Tamar and {Nakhleh}, Aidan J. and {Bharati Das}, Srijan},
        title = "{The Sun as a driver of the inner heliosphere: Modern investigations of fundamental plasma processes}",
      journal = {Nature Communications},
         year = 2026,
        month = may,
       volume = {17},
       number = {1},
          eid = {4184},
        pages = {4184},
          doi = {10.1038/s41467-026-72082-8},
       adsurl = {https://ui.adsabs.harvard.edu/abs/2026NatCo..17.4184E}
}

@ARTICLE{Howes-2024,
       author = {{Howes}, Gregory G.},
        title = "{The fundamental parameters of astrophysical plasma turbulence and its dissipation: non-relativistic limit}",
      journal = {Journal of Plasma Physics},
         year = 2024,
        month = oct,
       volume = {90},
       number = {5},
          eid = {905900504},
        pages = {905900504},
          doi = {10.1017/S0022377824001090},
archivePrefix = {arXiv},
       eprint = {2402.12829},
 primaryClass = {astro-ph.SR},
       adsurl = {https://ui.adsabs.harvard.edu/abs/2024JPlPh..90e9004H}
}

@phdthesis{Romeo-Thesis,
    author={Romeo,Orlando M.},
    year={2024},
    school={University of California, Berkeley},
    title={An Interdisciplinary Approach to Novel In Situ Measurements of Solar and Planetary Electrons and Magnetic Fields},
    journal={ProQuest Dissertations and Theses},
    pages={180},
    note={Copyright - Database copyright ProQuest LLC; ProQuest does not claim copyright in the individual underlying works; Last updated - 2025-04-24},
    isbn={9798310103153},
    language={English},
    url={https://www.proquest.com/dissertations-theses/interdisciplinary-approach-novel-situ/docview/3175813267/se-2},
    }

@ARTICLE{Chhiber-2019,
       author = {{Chhiber}, Rohit and {Usmanov}, Arcadi V. and {Matthaeus}, William H. and {Parashar}, Tulasi N. and {Goldstein}, Melvyn L.},
        title = "{Contextual Predictions for Parker Solar Probe. II. Turbulence Properties and Taylor Hypothesis}",
      journal = apjs,
         year = 2019,
        month = may,
       volume = {242},
       number = {1},
          eid = {12},
        pages = {12},
          doi = {10.3847/1538-4365/ab16d7},
archivePrefix = {arXiv},
       eprint = {1902.03340},
 primaryClass = {astro-ph.SR},
       adsurl = {https://ui.adsabs.harvard.edu/abs/2019ApJS..242...12C}
}

@ARTICLE{Kennel-1966,
       author = {{Kennel}, C.~F. and {Engelmann}, F.},
        title = "{Velocity Space Diffusion from Weak Plasma Turbulence in a Magnetic Field}",
      journal = {Physics of Fluids},
         year = 1966,
        month = dec,
       volume = {9},
       number = {12},
        pages = {2377-2388},
          doi = {10.1063/1.1761629},
       adsurl = {https://ui.adsabs.harvard.edu/abs/1966PhFl....9.2377K}
}

@ARTICLE{Klein-2016-FPC,
       author = {{Klein}, K.~G. and {Howes}, G.~G.},
        title = "{Measuring Collisionless Damping in Heliospheric Plasmas using Field-Particle Correlations}",
      journal = apjl,
         year = 2016,
        month = aug,
       volume = {826},
       number = {2},
          eid = {L30},
        pages = {L30},
          doi = {10.3847/2041-8205/826/2/L30},
archivePrefix = {arXiv},
       eprint = {1607.01738},
 primaryClass = {physics.space-ph},
       adsurl = {https://ui.adsabs.harvard.edu/abs/2016ApJ...826L..30K}
}

@ARTICLE{Cho-2009,
       author = {{Cho}, Jungyeon and {Lazarian}, A.},
        title = "{Simulations of Electron Magnetohydrodynamic Turbulence}",
      journal = apj,
         year = 2009,
        month = aug,
       volume = {701},
       number = {1},
        pages = {236-252},
          doi = {10.1088/0004-637X/701/1/236},
archivePrefix = {arXiv},
       eprint = {0904.0661},
 primaryClass = {astro-ph.EP},
       adsurl = {https://ui.adsabs.harvard.edu/abs/2009ApJ...701..236C}
}

@ARTICLE{Shankarappa-2024,
       author = {{Shankarappa}, Niranjana and {Klein}, Kristopher G. and {Martinovi{\'c}}, Mihailo M. and {Bowen}, Trevor A.},
        title = "{Estimated Heating Rates Due to Cyclotron Damping of Ion-scale Waves Observed by the Parker Solar Probe}",
      journal = apj,
         year = 2024,
        month = sep,
       volume = {973},
       number = {1},
          eid = {20},
        pages = {20},
          doi = {10.3847/1538-4357/ad5f2a},
archivePrefix = {arXiv},
       eprint = {2407.02708},
 primaryClass = {astro-ph.SR},
       adsurl = {https://ui.adsabs.harvard.edu/abs/2024ApJ...973...20S}
}

@ARTICLE{Zank-2022,
       author = {{Zank}, G.~P. and {Zhao}, L.-L. and {Adhikari}, L. and {Telloni}, D. and {Kasper}, J.~C. and {Stevens}, M. and {Rahmati}, A. and {Bale}, S.~D.},
        title = "{Turbulence in the Sub-Alfv{\'e}nic Solar Wind}",
      journal = apjl,
         year = 2022,
        month = feb,
       volume = {926},
       number = {2},
          eid = {L16},
        pages = {L16},
          doi = {10.3847/2041-8213/ac51da},
archivePrefix = {arXiv},
       eprint = {2202.02563},
 primaryClass = {astro-ph.SR},
       adsurl = {https://ui.adsabs.harvard.edu/abs/2022ApJ...926L..16Z}
}

@ARTICLE{Zhao-2022,
       author = {{Zhao}, L.-L. and {Zank}, G.~P. and {Telloni}, D. and {Stevens}, M. and {Kasper}, J.~C. and {Bale}, S.~D.},
        title = "{The Turbulent Properties of the Sub-Alfv{\'e}nic Solar Wind Measured by the Parker Solar Probe}",
      journal = apjl,
         year = 2022,
        month = apr,
       volume = {928},
       number = {2},
          eid = {L15},
        pages = {L15},
          doi = {10.3847/2041-8213/ac5fb0},
       adsurl = {https://ui.adsabs.harvard.edu/abs/2022ApJ...928L..15Z}
}

@ARTICLE{Zhang-2025,
       author = {{Zhang}, Michael F. and {Kunz}, Matthew W. and {Squire}, Jonathan and {Klein}, Kristopher G.},
        title = "{Extreme Heating of Minor Ions in Imbalanced Solar-wind Turbulence}",
      journal = apj,
         year = 2025,
        month = feb,
       volume = {979},
       number = {2},
          eid = {121},
        pages = {121},
          doi = {10.3847/1538-4357/ad95fc},
archivePrefix = {arXiv},
       eprint = {2408.04703},
 primaryClass = {astro-ph.SR},
       adsurl = {https://ui.adsabs.harvard.edu/abs/2025ApJ...979..121Z}
}

@ARTICLE{Meyrand-2021,
       author = {{Meyrand}, R. and {Squire}, J. and {Schekochihin}, A.~A. and {Dorland}, W.},
        title = "{On the violation of the zeroth law of turbulence in space plasmas}",
      journal = {Journal of Plasma Physics},
         year = 2021,
        month = may,
       volume = {87},
       number = {3},
          eid = {535870301},
        pages = {535870301},
          doi = {10.1017/S0022377821000489},
archivePrefix = {arXiv},
       eprint = {2009.02828},
 primaryClass = {physics.space-ph},
       adsurl = {https://ui.adsabs.harvard.edu/abs/2021JPlPh..87c5301M}
}

@ARTICLE{Squire-2022,
       author = {{Squire}, Jonathan and {Meyrand}, Romain and {Kunz}, Matthew W. and {Arzamasskiy}, Lev and {Schekochihin}, Alexander A. and {Quataert}, Eliot},
        title = "{High-frequency heating of the solar wind triggered by low-frequency turbulence}",
      journal = {Nature Astronomy},
         year = 2022,
        month = jun,
       volume = {6},
        pages = {715-723},
          doi = {10.1038/s41550-022-01624-z},
archivePrefix = {arXiv},
       eprint = {2109.03255},
 primaryClass = {astro-ph.SR},
       adsurl = {https://ui.adsabs.harvard.edu/abs/2022NatAs...6..715S}
}

@ARTICLE{Martinovic-2019,
       author = {{Martinovi{\'c}}, Mihailo M. and {Klein}, Kristopher G. and {Bourouaine}, Sofiane},
        title = "{Radial Evolution of Stochastic Heating in Low-{\ensuremath{\beta}} Solar Wind}",
      journal = apj,
         year = 2019,
        month = jul,
       volume = {879},
       number = {1},
          eid = {43},
        pages = {43},
          doi = {10.3847/1538-4357/ab23f4},
archivePrefix = {arXiv},
       eprint = {1905.13355},
 primaryClass = {physics.space-ph},
       adsurl = {https://ui.adsabs.harvard.edu/abs/2019ApJ...879...43M}
}

@ARTICLE{Klein-2015,
       author = {{Klein}, Kristopher G. and {Perez}, Jean C. and {Verscharen}, Daniel and {Mallet}, Alfred and {Chandran}, Benjamin D.~G.},
        title = "{A Modified Version of Taylor{\textquoteright}s Hypothesis for Solar Probe Plus Observations}",
      journal = apjl,
         year = 2015,
        month = mar,
       volume = {801},
       number = {1},
          eid = {L18},
        pages = {L18},
          doi = {10.1088/2041-8205/801/1/L18},
archivePrefix = {arXiv},
       eprint = {1412.3786},
 primaryClass = {physics.space-ph},
       adsurl = {https://ui.adsabs.harvard.edu/abs/2015ApJ...801L..18K}
}

@ARTICLE{Martinovic-2020,
       author = {{Martinovi{\'c}}, Mihailo M. and {Klein}, Kristopher G. and {Kasper}, Justin C. and {Case}, Anthony W. and {Korreck}, Kelly E. and {Larson}, Davin and {Livi}, Roberto and {Stevens}, Michael and {Whittlesey}, Phyllis and {Chandran}, Benjamin D.~G. and {Alterman}, Ben L. and {Huang}, Jia and {Chen}, Christopher H.~K. and {Bale}, Stuart D. and {Pulupa}, Marc and {Malaspina}, David M. and {Bonnell}, John W. and {Harvey}, Peter R. and {Goetz}, Keith and {Dudok de Wit}, Thierry and {MacDowall}, Robert J.},
        title = "{The Enhancement of Proton Stochastic Heating in the Near-Sun Solar Wind}",
      journal = apjs,
         year = 2020,
        month = feb,
       volume = {246},
       number = {2},
          eid = {30},
        pages = {30},
          doi = {10.3847/1538-4365/ab527f},
archivePrefix = {arXiv},
       eprint = {1912.02653},
 primaryClass = {astro-ph.SR},
       adsurl = {https://ui.adsabs.harvard.edu/abs/2020ApJS..246...30M}
}

@ARTICLE{Mallet-2019,
       author = {{Mallet}, Alfred and {Klein}, Kristopher G. and {Chandran}, Benjamin D.~G. and {Gro{\v{s}}elj}, Daniel and {Hoppock}, Ian W. and {Bowen}, Trevor A. and {Salem}, Chadi S. and {Bale}, Stuart D.},
        title = "{Interplay between intermittency and dissipation in collisionless plasma turbulence}",
      journal = {Journal of Plasma Physics},
         year = 2019,
        month = jun,
       volume = {85},
       number = {3},
          eid = {175850302},
        pages = {175850302},
          doi = {10.1017/S0022377819000357},
archivePrefix = {arXiv},
       eprint = {1807.09301},
 primaryClass = {physics.plasm-ph},
       adsurl = {https://ui.adsabs.harvard.edu/abs/2019JPlPh..85c1702M}
}

@ARTICLE{Vech-2018,
       author = {{Vech}, Daniel and {Mallet}, Alfred and {Klein}, Kristopher G. and {Kasper}, Justin C.},
        title = "{Magnetic Reconnection May Control the Ion-scale Spectral Break of Solar Wind Turbulence}",
      journal = apjl,
         year = 2018,
        month = mar,
       volume = {855},
       number = {2},
          eid = {L27},
        pages = {L27},
          doi = {10.3847/2041-8213/aab351},
archivePrefix = {arXiv},
       eprint = {1803.00065},
 primaryClass = {physics.space-ph},
       adsurl = {https://ui.adsabs.harvard.edu/abs/2018ApJ...855L..27V}
}

@ARTICLE{Verscharen-2019,
       author = {{Verscharen}, Daniel and {Klein}, Kristopher G. and {Maruca}, Bennett A.},
        title = "{The multi-scale nature of the solar wind}",
      journal = {Living Reviews in Solar Physics},
         year = 2019,
        month = dec,
       volume = {16},
       number = {1},
          eid = {5},
        pages = {5},
          doi = {10.1007/s41116-019-0021-0},
archivePrefix = {arXiv},
       eprint = {1902.03448},
 primaryClass = {physics.space-ph},
       adsurl = {https://ui.adsabs.harvard.edu/abs/2019LRSP...16....5V}
}

@ARTICLE{Bowen-2025prl,
       author = {{Bowen}, Trevor A. and {Ervin}, Tamar and {Mallet}, Alfred and {Chandran}, Benjamin D.~G. and {Sioulas}, Nikos and {Isenberg}, Philip A. and {Bale}, Stuart D. and {Squire}, Jonathan and {Klein}, Kristopher G. and {Pezzi}, Oreste},
        title = "{Stochastic Heating in the Sub-Alfv{\'e}nic Solar Wind}",
      journal = {\prl},
         year = 2025,
        month = dec,
       volume = {135},
       number = {25},
          eid = {255201},
        pages = {255201},
          doi = {10.1103/rxd8-22m9},
archivePrefix = {arXiv},
       eprint = {2509.20654},
 primaryClass = {astro-ph.SR},
       adsurl = {https://ui.adsabs.harvard.edu/abs/2025PhRvL.135y5201B}
}

@ARTICLE{Bowen-2024apjl,
       author = {{Bowen}, Trevor A. and {Vasko}, Ivan Y. and {Bale}, Stuart D. and {Chandran}, Benjamin D.~G. and {Chasapis}, Alexandros and {Dudok de Wit}, Thierry and {Mallet}, Alfred and {McManus}, Michael and {Meyrand}, Romain and {Pulupa}, Marc and {Squire}, Jonathan},
        title = "{Extended Cyclotron Resonant Heating of the Turbulent Solar Wind}",
      journal = apjl,
         year = 2024,
        month = sep,
       volume = {972},
       number = {1},
          eid = {L8},
        pages = {L8},
          doi = {10.3847/2041-8213/ad6b2e},
archivePrefix = {arXiv},
       eprint = {2406.10446},
 primaryClass = {astro-ph.SR},
       adsurl = {https://ui.adsabs.harvard.edu/abs/2024ApJ...972L...8B}
}

@ARTICLE{Isenberg-2007,
       author = {{Isenberg}, Philip A. and {Vasquez}, Bernard J.},
        title = "{Preferential Perpendicular Heating of Coronal Hole Minor Ions by the Fermi Mechanism}",
      journal = apj,
         year = 2007,
        month = oct,
       volume = {668},
       number = {1},
        pages = {546-556},
          doi = {10.1086/521220},
       adsurl = {https://ui.adsabs.harvard.edu/abs/2007ApJ...668..546I}
}

@ARTICLE{Mallet-2025SH,
       author = {{Mallet}, Alfred and {Klein}, Kristopher G. and {Chandran}, Benjamin D.~G. and {Ervin}, Tamar and {Bowen}, Trevor A.},
        title = "{Perpendicular ion heating in turbulence and reconnection: magnetic moment breaking by coherent fluctuations}",
      journal = {arXiv e-prints},
         year = 2025,
        month = sep,
          eid = {arXiv:2509.05518},
        pages = {arXiv:2509.05518},
          doi = {10.48550/arXiv.2509.05518},
archivePrefix = {arXiv},
       eprint = {2509.05518},
 primaryClass = {physics.plasm-ph},
       adsurl = {https://ui.adsabs.harvard.edu/abs/2025arXiv250905518M}
}

@article{Richardson-1955-finite, title={An Introduction to the Calculus of Finite Differences. By C.H. Richardson Pp. vi, 142. 28s. 1954. (Van Nostrand, New York; Macmillan, London)}, volume={39}, DOI={10.2307/3608616}, number={330}, journal={The Mathematical Gazette}, author={T.A.A.B.}, year={1955}, pages={339–339}}

@article{Wole-2021,
doi = {10.1088/1742-6596/2017/1/012009},
url = {https://dx.doi.org/10.1088/1742-6596/2017/1/012009},
year = {2021},
month = {sep},
publisher = {IOP Publishing},
volume = {2017},
number = {1},
pages = {012009},
author = {Wole, A and Lobo, M and Ginting, K Br.},
title = {The application of finite difference method on 2-D heat conductivity problem},
journal = {Journal of Physics: Conference Series}
}

@Inbook{Zhou-1993,
author="Zhou, Pei-bai",
title="Finite Difference Method",
bookTitle="Numerical Analysis of Electromagnetic Fields",
year="1993",
publisher="Springer Berlin Heidelberg",
address="Berlin, Heidelberg",
pages="63--94",
isbn="978-3-642-50319-1",
doi="10.1007/978-3-642-50319-1_3",
url="https://doi.org/10.1007/978-3-642-50319-1_3"
}

@INPROCEEDINGS{Bowen-2023-ursi,
       author = {{Bowen}, T.~A. and {Chandran}, B.~D.~G. and {Klein}, K.~G. and {Mallet}, A. and {Bale}, S.~D. and {Squire}, J. and {Verniero}, J.},
        title = "{Data-Driven Representations of Ion-Kinetic Distribution Functions}",
    booktitle = {2023 XXXVth General Assembly and Scientific Symposium of the International Union of Radio Science (URSI GASS)},
         year = 2023,
        month = oct,
          eid = {335},
        pages = {335},
          doi = {10.23919/URSIGASS57860.2023.10265538},
       adsurl = {https://ui.adsabs.harvard.edu/abs/2023ursi.confE.335B}
}

@Inbook{Branham-1990,
author="Branham, Richard L.",
title="The Singular Value Decomposition",
bookTitle="Scientific Data Analysis: An Introduction to Overdetermined Systems",
year="1990",
publisher="Springer New York",
address="New York, NY",
pages="199--232",
isbn="978-1-4612-3362-6",
doi="10.1007/978-1-4612-3362-6_8",
url="https://doi.org/10.1007/978-1-4612-3362-6_8"
}

@ARTICLE{Isenberg-1997,
       author = {{Isenberg}, Philip A.},
        title = "{A hemispherical model of anisotropic interstellar pickup ions}",
      journal = jgr,
         year = 1997,
        month = mar,
       volume = {102},
       number = {A3},
        pages = {4719-4724},
          doi = {10.1029/96JA03671},
       adsurl = {https://ui.adsabs.harvard.edu/abs/1997JGR...102.4719I}
}

@INPROCEEDINGS{Kulsrud-1983,
       author = {{Kulsrud}, R.~M.},
        title = "{MHD description of plasma}",
    booktitle = {Basic Plasma Physics: Selected Chapters, Handbook of Plasma Physics, Volume 1},
         year = 1983,
       editor = {{Galeev}, A.~A. and {Sudan}, R.~N.},
        month = jan,
        pages = {1},
       adsurl = {https://ui.adsabs.harvard.edu/abs/1983bpp..conf....1K}
}

@ARTICLE{Kasper-2021,
       author = {{Kasper}, J.~C. and {Klein}, K.~G. and {Lichko}, E. and {Huang}, Jia and {Chen}, C.~H.~K. and {Badman}, S.~T. and {Bonnell}, J. and {Whittlesey}, P.~L. and {Livi}, R. and {Larson}, D. and {Pulupa}, M. and {Rahmati}, A. and {Stansby}, D. and {Korreck}, K.~E. and {Stevens}, M. and {Case}, A.~W. and {Bale}, S.~D. and {Maksimovic}, M. and {Moncuquet}, M. and {Goetz}, K. and {Halekas}, J.~S. and {Malaspina}, D. and {Raouafi}, Nour E. and {Szabo}, A. and {MacDowall}, R. and {Velli}, Marco and {Dudok de Wit}, Thierry and {Zank}, G.~P.},
        title = "{Parker Solar Probe Enters the Magnetically Dominated Solar Corona}",
      journal = {\prl},
         year = 2021,
        month = dec,
       volume = {127},
       number = {25},
          eid = {255101},
        pages = {255101},
          doi = {10.1103/PhysRevLett.127.255101},
       adsurl = {https://ui.adsabs.harvard.edu/abs/2021PhRvL.127y5101K}
}

@ARTICLE{Cranmner-2000,
       author = {{Cranmer}, Steven R.},
        title = "{Ion Cyclotron Wave Dissipation in the Solar Corona: The Summed Effect of More than 2000 Ion Species}",
      journal = apj,
         year = 2000,
        month = apr,
       volume = {532},
       number = {2},
        pages = {1197-1208},
          doi = {10.1086/308620},
       adsurl = {https://ui.adsabs.harvard.edu/abs/2000ApJ...532.1197C}
}

@article{Johnston-2025,
  title = {Unified Phenomenology and Test-Particle Simulations of Ion Heating in Low-$\ensuremath{\beta}$ Plasmas},
  author = {Johnston, Zade and Squire, Jonathan and Meyrand, Romain},
  journal = {Phys. Rev. Lett.},
  volume = {135},
  issue = {9},
  pages = {095201},
  numpages = {9},
  year = {2025},
  month = {Aug},
  publisher = {American Physical Society},
  doi = {10.1103/rp85-hhg4},
  url = {https://link.aps.org/doi/10.1103/rp85-hhg4}
}

@ARTICLE{Chew-1956,
       author = {{Chew}, G.~F. and {Goldberger}, M.~L. and {Low}, F.~E.},
        title = "{The Boltzmann Equation and the One-Fluid Hydromagnetic Equations in the Absence of Particle Collisions}",
      journal = {Proceedings of the Royal Society of London Series A},
         year = 1956,
        month = jul,
       volume = {236},
       number = {1204},
        pages = {112-118},
          doi = {10.1098/rspa.1956.0116},
       adsurl = {https://ui.adsabs.harvard.edu/abs/1956RSPSA.236..112C}
}

@ARTICLE{Matteini-2007,
       author = {{Matteini}, Lorenzo and {Landi}, Simone and {Hellinger}, Petr and {Pantellini}, Filippo and {Maksimovic}, Milan and {Velli}, Marco and {Goldstein}, Bruce E. and {Marsch}, Eckart},
        title = "{Evolution of the solar wind proton temperature anisotropy from 0.3 to 2.5 AU}",
      journal = grl,
         year = 2007,
        month = oct,
       volume = {34},
       number = {20},
          eid = {L20105},
        pages = {L20105},
          doi = {10.1029/2007GL030920},
       adsurl = {https://ui.adsabs.harvard.edu/abs/2007GeoRL..3420105M}
}

@ARTICLE{Richardson-1995,
       author = {{Richardson}, John D. and {Paularena}, Karolen I. and {Lazarus}, Alan J. and {Belcher}, John W.},
        title = "{Radial evolution of the solar wind from IMP 8 to Voyager 2}",
      journal = grl,
         year = 1995,
        month = feb,
       volume = {22},
       number = {4},
        pages = {325-328},
          doi = {10.1029/94GL03273},
       adsurl = {https://ui.adsabs.harvard.edu/abs/1995GeoRL..22..325R}
}

@ARTICLE{Ervin-2024c,
       author = {{Ervin}, Tamar and {Jaffarove}, Kai and {Badman}, Samuel T. and {Huang}, Jia and {Rivera}, Yeimy J. and {Bale}, Stuart D.},
        title = "{Characteristics and Source Regions of Slow Alfv{\'e}nic Solar Wind Observed by Parker Solar Probe}",
      journal = apj,
         year = 2024,
        month = nov,
       volume = {975},
       number = {2},
          eid = {156},
        pages = {156},
          doi = {10.3847/1538-4357/ad7d00},
archivePrefix = {arXiv},
       eprint = {2407.09684},
 primaryClass = {astro-ph.SR},
       adsurl = {https://ui.adsabs.harvard.edu/abs/2024ApJ...975..156E}
}

@ARTICLE{Huang-2025,
       author = {{Huang}, Jia and {Larson}, Davin E. and {Ervin}, Tamar and {Liu}, Mingzhe and {Ortiz}, Oscar and {Martinovi{\'c}}, Mihailo M. and {Huang}, Zhenguang and {Chasapis}, Alexandros and {Chu}, Xiangning and {Alterman}, B.~L. and {Huang}, Zesen and {Wei}, Wenwen and {Verniero}, J.~L. and {Jian}, Lan K. and {Szabo}, Adam and {Romeo}, Orlando and {Rahmati}, Ali and {Livi}, Roberto and {Whittlesey}, Phyllis and {Alnussirat}, Samer T. and {Kasper}, Justin C. and {Stevens}, Michael and {Bale}, Stuart D.},
        title = "{The Temperature Anisotropy and Helium Abundance Features of Alfv{\'e}nic Slow Solar Wind Observed by Parker Solar Probe, Helios, and Wind Missions}",
      journal = apjl,
         year = 2025,
        month = jun,
       volume = {986},
       number = {2},
          eid = {L28},
        pages = {L28},
          doi = {10.3847/2041-8213/ade0ac},
archivePrefix = {arXiv},
       eprint = {2005.12372},
 primaryClass = {physics.space-ph},
       adsurl = {https://ui.adsabs.harvard.edu/abs/2025ApJ...986L..28H}
}

@ARTICLE{Marsch-1982VDF,
       author = {{Marsch}, E. and {Schwenn}, R. and {Rosenbauer}, H. and {Muehlhaeuser}, K. -H. and {Pilipp}, W. and {Neubauer}, F.~M.},
        title = "{Solar wind protons: Three-dimensional velocity distributions and derived plasma parameters measured between 0.3 and 1 AU}",
      journal = jgr,
         year = 1982,
        month = jan,
       volume = {87},
       number = {A1},
        pages = {52-72},
          doi = {10.1029/JA087iA01p00052},
       adsurl = {https://ui.adsabs.harvard.edu/abs/1982JGR....87...52M}
}

@article{Bowen-2020-SCaM,
	adsurl = {https://ui.adsabs.harvard.edu/abs/2020JGRA..12527813B},
	archiveprefix = {arXiv},
	author = {{Bowen}, T.~A. and {Bale}, S.~D. and {Bonnell}, J.~W. and {Dudok de Wit}, T. and {Goetz}, K. and {Goodrich}, K. and {Gruesbeck}, J. and {Harvey}, P.~R. and {Jannet}, G. and {Koval}, A. and {MacDowall}, R.~J. and {Malaspina}, D.~M. and {Pulupa}, M. and {Revillet}, C. and {Sheppard}, D. and {Szabo}, A.},
	doi = {10.1029/2020JA027813},
	eid = {e27813},
	eprint = {2001.04587},
	journal = {Journal of Geophysical Research (Space Physics)},
	month = may,
	number = {5},
	pages = {e27813},
	primaryclass = {astro-ph.IM},
	title = {{A Merged Search-Coil and Fluxgate Magnetometer Data Product for Parker Solar Probe FIELDS}},
	volume = {125},
	year = 2020}

@article{Kasper-2016,
	author = {Kasper, Justin C. and Abiad, Robert and Austin, Gerry and Balat-Pichelin, Marianne and Bale, Stuart D. and Belcher, John W. and Berg, Peter and Bergner, Henry and Berthomier, Matthieu and Bookbinder, Jay and Brodu, Etienne and Caldwell, David and Case, Anthony W. and Chandran, Benjamin D. G. and Cheimets, Peter and Cirtain, Jonathan W. and Cranmer, Steven R. and Curtis, David W. and Daigneau, Peter and Dalton, Greg and Dasgupta, Brahmananda and DeTomaso, David and Diaz-Aguado, Millan and Djordjevic, Blagoje and Donaskowski, Bill and Effinger, Michael and Florinski, Vladimir and Fox, Nichola and Freeman, Mark and Gallagher, Dennis and Gary, S. Peter and Gauron, Tom and Gates, Richard and Goldstein, Melvin and Golub, Leon and Gordon, Dorothy A. and Gurnee, Reid and Guth, Giora and Halekas, Jasper and Hatch, Ken and Heerikuisen, Jacob and Ho, George and Hu, Qiang and Johnson, Greg and Jordan, Steven P. and Korreck, Kelly E. and Larson, Davin and Lazarus, Alan J. and Li, Gang and Livi, Roberto and Ludlam, Michael and Maksimovic, Milan and McFadden, James P. and Marchant, William and Maruca, Bennet A. and McComas, David J. and Messina, Luciana and Mercer, Tony and Park, Sang and Peddie, Andrew M. and Pogorelov, Nikolai and Reinhart, Matthew J. and Richardson, John D. and Robinson, Miles and Rosen, Irene and Skoug, Ruth M. and Slagle, Amanda and Steinberg, John T. and Stevens, Michael L. and Szabo, Adam and Taylor, Ellen R. and Tiu, Chris and Turin, Paul and Velli, Marco and Webb, Gary and Whittlesey, Phyllis and Wright, Ken and Wu, S. T. and Zank, Gary},
	date = {2016/12/01},
	doi = {10.1007/s11214-015-0206-3},
	id = {Kasper2016},
	isbn = {1572-9672},
	journal = {Space Science Reviews},
	number = {1},
	pages = {131--186},
	title = {Solar Wind Electrons Alphas and Protons (SWEAP) Investigation: Design of the Solar Wind and Coronal Plasma Instrument Suite for Solar Probe Plus},
	url = {https://doi.org/10.1007/s11214-015-0206-3},
	volume = {204},
	year = {2016}}

@article{Isenberg-2019,
	adsurl = {https://ui.adsabs.harvard.edu/abs/2019ApJ...887...63I},
	author = {{Isenberg}, Philip A. and {Vasquez}, Bernard J.},
	doi = {10.3847/1538-4357/ab4e12},
	eid = {63},
	journal = apj,
	month = dec,
	number = {1},
	pages = {63},
	title = {{Perpendicular Ion Heating by Cyclotron Resonant Dissipation of Turbulently Generated Kinetic Alfv{\'e}n Waves in the Solar Wind}},
	volume = {887},
	year = 2019}

@article{Vasquez-2020,
	adsurl = {https://ui.adsabs.harvard.edu/abs/2020ApJ...893...71V},
	author = {{Vasquez}, Bernard J. and {Isenberg}, Philip A. and {Markovskii}, Sergei A.},
	doi = {10.3847/1538-4357/ab7e2b},
	eid = {71},
	journal = apj,
	month = apr,
	number = {1},
	pages = {71},
	title = {{Proton Perpendicular Heating in Turbulence Simulations: Determination of the Velocity Diffusion Coefficient}},
	volume = {893},
	year = 2020}

@article{Klein-2016,
	adsurl = {https://ui.adsabs.harvard.edu/abs/2016ApJ...820...47K},
	archiveprefix = {arXiv},
	author = {{Klein}, Kristopher G. and {Chandran}, Benjamin D.~G.},
	doi = {10.3847/0004-637X/820/1/47},
	eid = {47},
	eprint = {1602.05114},
	journal = apj,
	month = mar,
	number = {1},
	pages = {47},
	primaryclass = {astro-ph.SR},
	title = {{Evolution of The Proton Velocity Distribution due to Stochastic Heating in the Near-Sun Solar Wind}},
	volume = {820},
	year = 2016}

@article{Bourouaine-2013,
	adsurl = {https://ui.adsabs.harvard.edu/abs/2013ApJ...774...96B},
	archiveprefix = {arXiv},
	author = {{Bourouaine}, Sofiane and {Chandran}, Benjamin D.~G.},
	doi = {10.1088/0004-637X/774/2/96},
	eid = {96},
	eprint = {1307.3789},
	journal = apj,
	month = sep,
	number = {2},
	pages = {96},
	primaryclass = {astro-ph.SR},
	title = {{Observational Test of Stochastic Heating in Low-{\ensuremath{\beta}} Fast-solar-wind Streams}},
	volume = {774},
	year = 2013}

@article{Chandran-2010,
	adsurl = {https://ui.adsabs.harvard.edu/abs/2010ApJ...720..548C},
	archiveprefix = {arXiv},
	author = {{Chandran}, Benjamin D.~G.},
	doi = {10.1088/0004-637X/720/1/548},
	eprint = {1006.3473},
	journal = apj,
	month = sep,
	number = {1},
	pages = {548-554},
	primaryclass = {astro-ph.SR},
	title = {{Alfv{\'e}n-wave Turbulence and Perpendicular Ion Temperatures in Coronal Holes}},
	volume = {720},
	year = 2010}

@article{Bowen-2024,
	adsurl = {https://ui.adsabs.harvard.edu/abs/2024NatAs...8..482B},
	archiveprefix = {arXiv},
	author = {{Bowen}, Trevor A. and {Bale}, Stuart D. and {Chandran}, Benjamin D.~G. and {Chasapis}, Alexandros and {Chen}, Christopher H.~K. and {Dudok de Wit}, Thierry and {Mallet}, Alfred and {Meyrand}, Romain and {Squire}, Jonathan},
	doi = {10.1038/s41550-023-02186-4},
	eprint = {2306.04881},
	journal = {Nature Astronomy},
	month = apr,
	pages = {482-490},
	primaryclass = {physics.space-ph},
	title = {{Mediation of collisionless turbulent dissipation through cyclotron resonance}},
	volume = {8},
	year = 2024}

@article{Cerri-2021,
	adsurl = {https://ui.adsabs.harvard.edu/abs/2021ApJ...916..120C},
	archiveprefix = {arXiv},
	author = {{Cerri}, S.~S. and {Arzamasskiy}, L. and {Kunz}, M.~W.},
	doi = {10.3847/1538-4357/abfbde},
	eid = {120},
	eprint = {2102.09654},
	journal = apj,
	month = aug,
	number = {2},
	pages = {120},
	primaryclass = {astro-ph.SR},
	title = {{On Stochastic Heating and Its Phase-space Signatures in Low-beta Kinetic Turbulence}},
	volume = {916},
	year = 2021}

@article{Arzamasskiy-2019,
	adsurl = {https://ui.adsabs.harvard.edu/abs/2019ApJ...879...53A},
	archiveprefix = {arXiv},
	author = {{Arzamasskiy}, Lev and {Kunz}, Matthew W. and {Chandran}, Benjamin D.~G. and {Quataert}, Eliot},
	doi = {10.3847/1538-4357/ab20cc},
	eid = {53},
	eprint = {1901.11028},
	journal = apj,
	month = jul,
	number = {1},
	pages = {53},
	primaryclass = {astro-ph.HE},
	title = {{Hybrid-kinetic Simulations of Ion Heating in Alfv{\'e}nic Turbulence}},
	volume = {879},
	year = 2019}

@article{Bowen-2022,
	adsurl = {https://ui.adsabs.harvard.edu/abs/2022PhRvL.129p5101B},
	archiveprefix = {arXiv},
	author = {{Bowen}, Trevor A. and {Chandran}, Benjamin D.~G. and {Squire}, Jonathan and {Bale}, Stuart D. and {Duan}, Die and {Klein}, Kristopher G. and {Larson}, Davin and {Mallet}, Alfred and {McManus}, Michael D. and {Meyrand}, Romain and {Verniero}, Jaye L. and {Woodham}, Lloyd D.},
	doi = {10.1103/PhysRevLett.129.165101},
	eid = {165101},
	eprint = {2111.05400},
	journal = {\prl},
	month = oct,
	number = {16},
	pages = {165101},
	primaryclass = {astro-ph.SR},
	title = {{In Situ Signature of Cyclotron Resonant Heating in the Solar Wind}},
	volume = {129},
	year = 2022}

@article{Fox-2016,
	adsurl = {https://ui.adsabs.harvard.edu/abs/2016SSRv..204....7F},
	author = {{Fox}, N.~J. and {Velli}, M.~C. and {Bale}, S.~D. and {Decker}, R. and {Driesman}, A. and {Howard}, R.~A. and {Kasper}, J.~C. and {Kinnison}, J. and {Kusterer}, M. and {Lario}, D. and {Lockwood}, M.~K. and {McComas}, D.~J. and {Raouafi}, N.~E. and {Szabo}, A.},
	doi = {10.1007/s11214-015-0211-6},
	journal = ssr,
	month = dec,
	number = {1-4},
	pages = {7-48},
	title = {{The Solar Probe Plus Mission: Humanity's First Visit to Our Star}},
	volume = {204},
	year = 2016}

@article{Livi-2022,
	author = {Roberto Livi and Davin E. Larson and Justin C. Kasper and Robert Abiad and A. W. Case and Kristopher G. Klein and David W. Curtis and Gregory Dalton and Michael Stevens and Kelly E. Korreck and George Ho and Miles Robinson and Chris Tiu and Phyllis L. Whittlesey and Jaye L. Verniero and Jasper Halekas and James McFadden and Mario Marckwordt and Amanda Slagle and Mamuda Abatcha and Ali Rahmati and Michael D. McManus},
	doi = {10.3847/1538-4357/ac93f5},
	journal = {The Astrophysical Journal},
	month = {oct},
	number = {2},
	pages = {138},
	publisher = {The American Astronomical Society},
	title = {The Solar Probe ANalyzer---Ions on the Parker Solar Probe},
	url = {https://dx.doi.org/10.3847/1538-4357/ac93f5},
	volume = {938},
	year = {2022}}

@article{Bale-2016,
	adsurl = {https://ui.adsabs.harvard.edu/abs/2016SSRv..204...49B},
	author = {{Bale}, S.~D. and {Goetz}, K. and {Harvey}, P.~R. and {Turin}, P. and {Bonnell}, J.~W. and {Dudok de Wit}, T. and {Ergun}, R.~E. and {MacDowall}, R.~J. and {Pulupa}, M. and {Andre}, M. and {Bolton}, M. and {Bougeret}, J. -L. and {Bowen}, T.~A. and {Burgess}, D. and {Cattell}, C.~A. and {Chandran}, B.~D.~G. and {Chaston}, C.~C. and {Chen}, C.~H.~K. and {Choi}, M.~K. and {Connerney}, J.~E. and {Cranmer}, S. and {Diaz-Aguado}, M. and {Donakowski}, W. and {Drake}, J.~F. and {Farrell}, W.~M. and {Fergeau}, P. and {Fermin}, J. and {Fischer}, J. and {Fox}, N. and {Glaser}, D. and {Goldstein}, M. and {Gordon}, D. and {Hanson}, E. and {Harris}, S.~E. and {Hayes}, L.~M. and {Hinze}, J.~J. and {Hollweg}, J.~V. and {Horbury}, T.~S. and {Howard}, R.~A. and {Hoxie}, V. and {Jannet}, G. and {Karlsson}, M. and {Kasper}, J.~C. and {Kellogg}, P.~J. and {Kien}, M. and {Klimchuk}, J.~A. and {Krasnoselskikh}, V.~V. and {Krucker}, S. and {Lynch}, J.~J. and {Maksimovic}, M. and {Malaspina}, D.~M. and {Marker}, S. and {Martin}, P. and {Martinez-Oliveros}, J. and {McCauley}, J. and {McComas}, D.~J. and {McDonald}, T. and {Meyer-Vernet}, N. and {Moncuquet}, M. and {Monson}, S.~J. and {Mozer}, F.~S. and {Murphy}, S.~D. and {Odom}, J. and {Oliverson}, R. and {Olson}, J. and {Parker}, E.~N. and {Pankow}, D. and {Phan}, T. and {Quataert}, E. and {Quinn}, T. and {Ruplin}, S.~W. and {Salem}, C. and {Seitz}, D. and {Sheppard}, D.~A. and {Siy}, A. and {Stevens}, K. and {Summers}, D. and {Szabo}, A. and {Timofeeva}, M. and {Vaivads}, A. and {Velli}, M. and {Yehle}, A. and {Werthimer}, D. and {Wygant}, J.~R.},
	doi = {10.1007/s11214-016-0244-5},
	journal = ssr,
	month = dec,
	number = {1-4},
	pages = {49-82},
	title = {{The FIELDS Instrument Suite for Solar Probe Plus. Measuring the Coronal Plasma and Magnetic Field, Plasma Waves and Turbulence, and Radio Signatures of Solar Transients}},
	volume = {204},
	year = 2016}

\section{End Matter} \label{sec:em}
\begin{figure}
    \centering
    \includegraphics[width=\linewidth]{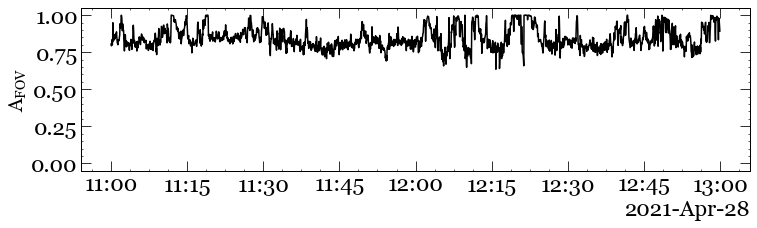}
    \caption{PSP/SPANi FOV metric combining energy flux vs. $\theta$ and $\phi$ over the period of interest as observed by PSP/SPANi \cite{Romeo-Thesis}.}
    \label{fig:fov}
\end{figure}

\paragraph{SPANi Field-of-View (FOV)}
In Fig.~\ref{fig:fov}, we show a quantification of the SPANi FOV that combines both look directions, $\theta$ and $\phi$ \citep{Romeo-Thesis}. The high metric throughout the interval indicates that the proton distribution function is well within the instruments FOV and thus suitable for this analysis.

\paragraph{Calculation of the Stochastic Heating Rate} \label{end:sh-calculation}

There are various formulations for analytic expressions for stochastic heating, specifically differences definitions of $\delta v$ and the associated stochasticity parameter ($\varepsilon = \delta v / v$). We outline the methods we compare below:

\begin{enumerate}
    \item In KC16 \citep{Klein-2016}, $\varepsilon = \delta v_\rho / v_\perp$ where $\delta v_\rho$ is the rms fluctuation amplitude at each scale $\rho = v_\perp / \Omega$, making $\varepsilon$ scale dependent.
    \item J25 \citep{Johnston-2025} define $\varepsilon = \hat{\xi_i} = {\duJS} / v_{th, i}$ where ${\duJS}$ are the velocity fluctuations at the perpendicular scale corresponding to the smallest nonlinear timescale. This can account for the helicity barrier, which causes the power spectrum to steepen at a wavenumber $k_\perp^*$ that is significantly smaller than $\rho_{th, p}^{-1}$.
    
\end{enumerate}



We use PSP SCaM (search coil and magnetometer) data which provides a two-axis measurement of the magnetic field at high fidelity to calculate $\delta v$ over the same 116 time intervals as for the empirical calculation. We point the reader to B25 for additional discussion of the assumptions. We note that the inversion method described in the main text calculates the velocity-space heating rate independent of these assumptions.
\begin{enumerate}
    \item $\delta \mathbf{B} (f, t)$: We calculate the scale dependent $\delta \mathbf{B}$ using five-point increments with 88 time lags ranging from 0.85 ms $< \; \tau \; <$ 7.23 s (0.07 $< \; f \; <$ 586 Hz) where $f = 1 / 2 \tau$ \citep{Cho-2009}. 
    
    \item $\delta v(k_\perp, t)$: We assume {\alfic} turbulence to convert $\delta B(f, t)$ to $\delta v(f, t)$ whereby $\delta v / v_{phase}^{KAW} = \delta B / B_0$ using $v_{phase}^{KAW} = v_A \sqrt{1 + k_\perp^2 \rho_{th}^2}$. This conversion ignores the $k_\parallel$ term (e.g. $k_\parallel = 0$) such that $k_\perp = 2 \pi f / (v_{SW} \sin{\theta_{VB}})$. We note this correction is valid in the frequency regime where $\omega < \Omega_p$ ($\Omega_p = q B / m_p$). 

    \item $\delta v(v_\perp, t)$: We use the Taylor hypothesis to determine corresponding physical scales. This allows us to relate each time lag ($\tau$) to wave wavenumbers $k_\perp$ and $k_\parallel$ assuming the wave vector is in the plane containing $v_{SW}$ and $\mathbf{B}$:
    \begin{equation} \label{eqn:doppler-shift}
        2 \pi f = \mathbf{k} \cdot \mathbf{V} = 
        k_\parallel v_A + k_\parallel v_{SW} \cos{\theta_{VB}} + k_\perp v_{SW} \sin{\theta_{VB}}
    \end{equation}
    We use the condition that $k_\perp \rho = 1$ giving $k_\perp = \Omega_i / v_\perp$ to determine $\delta v(v_\perp, t)$ following the Modified Taylor hypothesis (MTH) as relevant to flows where $v_{SW} < \sim v_A$ \cite{Klein-2015, Chhiber-2019}. We assume critical balance where $k_\parallel v_A = \delta v k_\perp = \delta v \Omega_p / v_\perp$, giving 
    \begin{multline}
        v_\perp^{MTH} = \frac{\Omega_p}{2 \pi f} \big[ \delta v  \left(1 + \frac{v_{SW}}{v_A} \cos{\theta_{VB}} \right) 
        \\
         +  v_{SW} \sin{\theta_{VB}}] .
    \end{multline}

    \item $\duJS(t)$: We identify the relevant perpendicular scale: $k_\perp^m$ which is where the non-linear growth rate, $k_\perp \delta v (k_\perp)$, is maximized. We determine $v_\perp^m$, assuming $k_\perp^m \rho = 1$ giving $v_\perp^m = \Omega_i / k_\perp^m$ leading to: $\duJS(t) =  \delta v (v_\perp^m, t)$ for calculation of the J25 coefficient. 
\end{enumerate}

Upon calculation of $\delta v$, we calculate the \lq{}rms\rq{}, \lq{}intermittent\rq{}, and J25 diffusion coefficients (Eqs.~\ref{eqn:coeff-rms-int}-\ref{eqn:coeff-J25}). $T$ is the length of the non-overlapping window used to calculate $\delta v$ from the measurements. The resulting heating rates calculated following Eq.~\ref{eqn:perp-heating} are shown in Fig.~\ref{fig:sh-comparison}.
\begin{align} \label{eqn:coeff-rms-int}
    \dperprms (v_\perp) = c_1 \frac{\Omega_i}{v_\perp} \langle \delta v(v_\perp, t) \rangle_T^3 e^{-c_2 v_\perp / \langle \delta v(v_\perp, t) \rangle_T} \\
    \dperpint (v_\perp) = \big \langle c_1 \frac{\Omega_i}{v_\perp} \delta v(v_\perp, t)^3 e^{-c_2 v_\perp / \delta v(v_\perp, t)} \big \rangle_T \\
    \dperpintJ = \big \langle c_1 \frac{\Omega_i}{v_\perp} \duJS(t)^3 e^{-c_2 v_\perp / \duJS(t)} \big \rangle_{T}  \label{eqn:coeff-J25}
\end{align}

\paragraph{Calculation of the Ion-Cyclotron Heating Rate} \label{end:icw-calculation}
We calculate a perpendicular ion-cyclotron heating rate for comparison with $\dQperpemp$. We take the quasilinear expression for the resonant cyclotron effect on the proton distribution due to a spectrum of parallel propagating ICWs \citep{Kennel-1966, Isenberg-2007, Bowen-2024apjl}, form the product of this with $\frac{1}{2} m v^2$, and integrate over phase space to obtain a quasilinear heating rate:

\begin{equation}
    Q^{\mathrm{ICW}} = \int \frac{1}{2} m_p v^2 \frac{d f}{dt} \mathrm{d^3}v .
\end{equation}

\begin{figure}
    \centering
    \includegraphics[width=\linewidth]{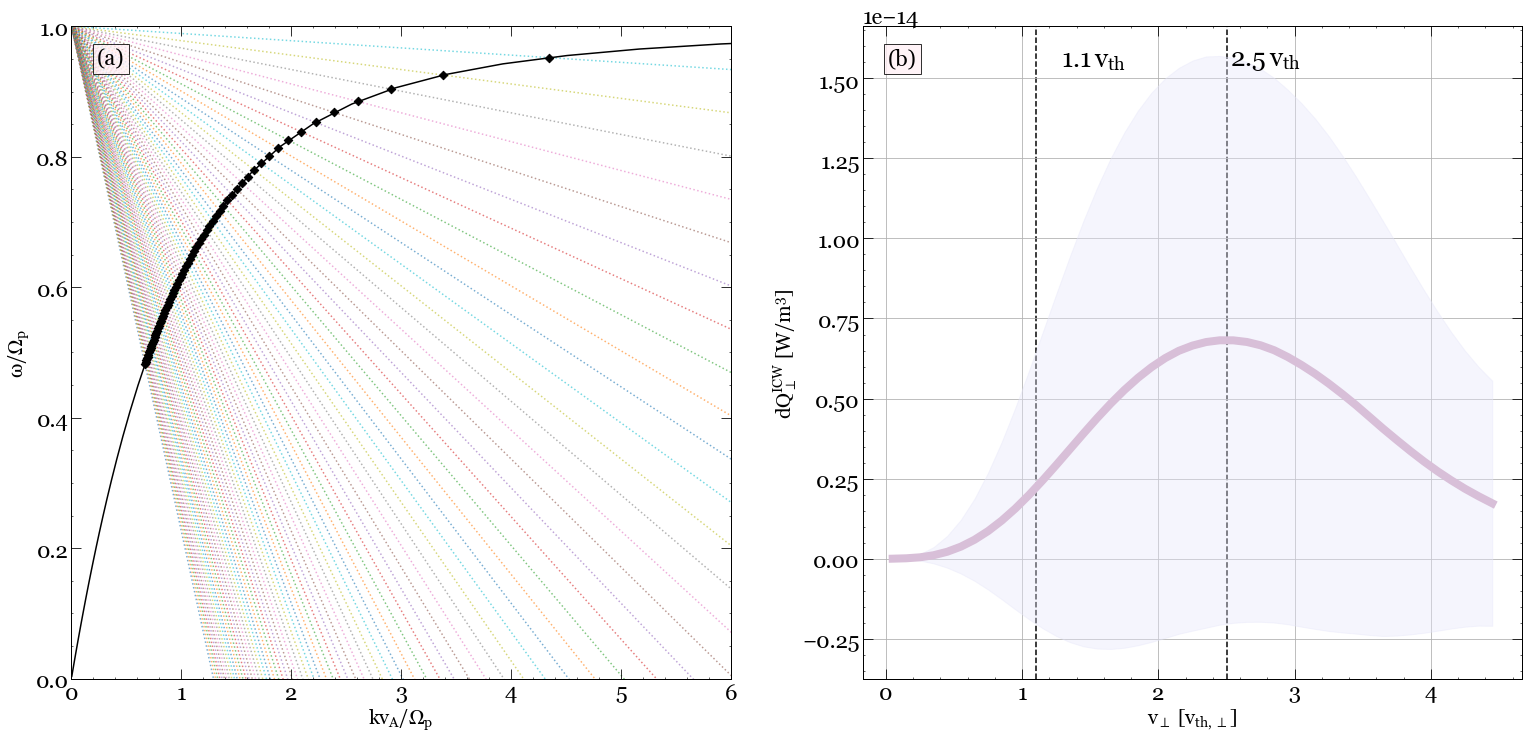}
    \caption{Panel (a): Representation of the cyclotron resonance condition used to determine $\kres$ and $\wres$ for calculation of the phase and group speeds. The solid black line shows the dispersion relation for parallel-propagating ICWs in a cold electron-proton plasma. The colored lines show the resonance condition $\omega(k) = k v_\parallel + \Omega_i$ for values of $v_\parallel$ in our grid (we note resonance only occurs for $v_\parallel <0$, which is what is shown here). The diamonds represent the points corresponding to $(\kres, \wres)$ for each $v_\parallel$. Panel (b): Average differential heating rate ($\dQperpicw$) calculated from Eq.~\ref{eqn:icw-heating} and $\parallel$-ICW spectra \citep{Bowen-2022, Bowen-2024apjl}. The dashed line at $2.5 v_{th, \perp}$ ($1.1 v_{th, \perp}$) shows the location of the peak in $\dQperpicw$ ($\dQperpemp$). The purple shaded region shows the error from the empirical calculation.
    }
    \label{fig:icw-res}
\end{figure}

After integration by parts we find the following:

\begin{multline}
\label{eqn:icw-heating-full}
    Q^{\mathrm{ICW}} = -\pi^2 \Omega_p^2 m_p
    \int dv_\perp \int dv_\parallel
    \int dk \, v_\perp^2 \left( \frac{\omega}{k} \right)^2 \\
    {}\times \delta\!\left( \omega - k v_\parallel - \Omega_p \right)
    I(k)
    \left[
    \left( 1 - \frac{k v_\parallel}{\omega} \right)
    \frac{\partial f}{\partial v_\perp}
    + \frac{k v_\perp}{\omega}
    \frac{\partial f}{\partial v_\parallel}
    \right] .
\end{multline}
Here the definition of the delta function imposes the resonance condition 
\begin{equation} \label{eqn:icw-res}
    \omega (k) = k v_\parallel  + \Omega_p
\end{equation}
as a function of $v_\parallel$ of the proton and the dispersion relation for $\parallel$-ICWs. The dispersion relation in a cold electron-proton plasma is
\begin{equation} \label{eqn:icw-disp}
    \frac{\omega}{k} = v_A \sqrt{1 - \frac{\omega}{\Omega_p}}
\end{equation}
where we assume only outward propagation. The proton resonance with these dispersive waves is illustrated in Fig.~\ref{fig:icw-res}(a), depicting
the one-to-one relationship between the proton $v_\parallel$ and the resonant wave. We define a $(v_\perp, v_\parallel)$ grid that corresponds to the uniformly spaced VDF grid (see Supplementary Material). For each $v_\parallel$, we define a resonant wavenumber ($\kres$) and frequency ($\wres$) based on Eq.~\ref{eqn:icw-res}. The diamonds in Fig.~\ref{fig:icw-res}(a) indicate the values of
$\wres$ and $\kres$ where the resonance condition and the dispersion relation are both satisfied on this grid. $\kres$ and $\wres$ are then used to determine the phase and group speeds of the resonant wave \citep{Isenberg-2007}:
\begin{align} \label{eqn:vph-vg}
    V_{res} = \frac{\wres}{\kres} \\
    W_{res} = \frac{2 V_{res}^2}{V_{res}^2 + v_A^2}.
\end{align}

Using the $\delta$-function to substitute in $\kres$, we define a differential heating rate where $Q^\mathrm{ICW} = \int d v_\perp \int  d v_\parallel \; dQ^{\mathrm{ICW}} $.
\begin{multline} \label{eqn:icw-heating}
        dQ^{\mathrm{ICW}} =
         - \pi^2 \Omega_p^2 m_p \\ \left\{
        V_{res} \frac{v_\perp^2 I(k_{res})}{|W_{res} - v_\parallel|} 
    \left [(V_{res} - v_\parallel)\frac{\partial f}{\partial v_\perp} + v_\perp \frac{\partial f}{\partial v_\parallel} \right] \right\}
\end{multline}
Here, $I(k_{res})$ is the ICW intensity associated with the resonance of proton at a specific $v_\parallel$. From PSP SCaM observations, we produce a spectrum $I(k)$ \citep{Bowen-2022, Bowen-2024apjl} at a 56 second cadence, producing 2064 spectra over the 2-hour interval. Each $I(k)$ is interpolated to produce $I(\kres)$ where $\kres$ are the resonant wave numbers corresponding to each $v_\parallel$.

We formulate a differential perpendicular ICW heating rate in a manner similar to the empirical and SH rates shown in Fig.~\ref{fig:sh-comparison}: $\dQperpicw = \int  d v_\parallel dQ^{\mathrm{ICW}} \Delta v_\perp$ shown in Fig.~\ref{fig:icw-res}(b). The $v_\perp$ and $v_\parallel$ grids used here are equivalent to the grid used for the interpolation of the VDFs and calculation of $\dQperpsh$. We find that the heating rate determined from the ICW calculation peaks at $\sim 2.5 v_{th}$, in comparison to the peak near $1.1 v_{th}$ in $\dQperpemp$ and $\dQperpint$, and is far too small to account for our observed heating (Fig.~\ref{fig:sh-comparison}).


\providecommand{\noopsort}[1]{}\providecommand{\singleletter}[1]{#1}

\end{document}